\documentclass[reprint,twocolumn,nofootinbib,prl,floatfix,superscriptaddress]{revtex4-1}
\usepackage[utf8]{inputenc}
\usepackage[T1]{fontenc}

\usepackage{amsmath}
\usepackage{amssymb}
\usepackage{braket}
\usepackage{siunitx}
\usepackage[version=4]{mhchem}
\usepackage{csquotes}

\usepackage{tikz}
\usetikzlibrary{fit,positioning,calc}

\usepackage[breaklinks=true,colorlinks=true,linkcolor=blue,urlcolor=blue,citecolor=blue]{hyperref}
\usepackage{bookmark}

\def\dd{\ensuremath{\mathrm{d}}}

\def\ii{\ensuremath{\mathrm{i}}}

\def\ZZ{\ensuremath{\mathbb{Z}}}

\def\tr{\ensuremath{\operatorname{tr}}}
\def\Tr{\ensuremath{\operatorname{Tr}}}

\def\Id{\mathbf{I}}

\def\strong#1{\textbf{#1}}

\def\paragraphTitle#1{\strong{#1}. }

\begin{document}
\title{Machine learning assisted measurement of local topological invariants}
\author{Marcello D. Caio}
\email{caio@lorentz.leidenuniv.nl}
\affiliation{Instituut-Lorentz, Universiteit Leiden, P.O. Box 9506, 2300 RA Leiden, The Netherlands}
\author{Marco Caccin}
\affiliation{Independent Researcher}
\author{Paul Baireuther}
\affiliation{Instituut-Lorentz, Universiteit Leiden, P.O. Box 9506, 2300 RA Leiden, The Netherlands}
\author{Timo Hyart}
\affiliation{International Research Centre MagTop, Institute of Physics, Polish Academy
of Sciences, Aleja Lotnikow 32/46, PL-02668 Warsaw, Poland}
\author{Michel Fruchart}
\email{fruchart@lorentz.leidenuniv.nl}
\affiliation{Instituut-Lorentz, Universiteit Leiden, P.O. Box 9506, 2300 RA Leiden, The Netherlands}
\affiliation{The James Franck Institute, The University of Chicago, Chicago, IL 60637, USA}

\makeatletter
\renewcommand\frontmatter@abstractwidth{\dimexpr\textwidth\relax}
\makeatother

\begin{abstract}
The continuous effort towards topological quantum devices calls for an efficient and non-invasive method to assess the conformity of components in different topological phases. Here, we show that machine learning paves the way towards non-invasive topological quality control. To do so, we use a local topological marker, able to discriminate between topological phases of one-dimensional wires. 
The direct observation of this marker in solid state systems is challenging, but we show that an artificial neural network can learn to approximate it from the experimentally accessible local density of states. Our method distinguishes different non-trivial phases, even for systems where direct transport measurements are not available and for composite systems. This new approach could find significant use in experiments, ranging from the study of novel topological materials to high-throughput automated material design.
\end{abstract}
\maketitle

\begin{figure}
	\includegraphics{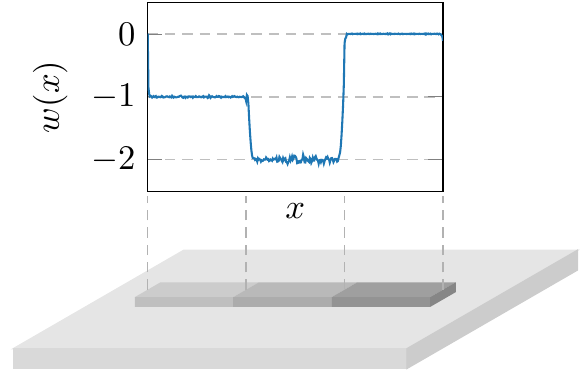}
	\caption{\label{fig_winding_marker}\strong{Winding marker in a composite sample.} We show a hypothetical one-dimensional quantum device composed of three regions in different topological phases. These are distinguished by quantised values of the winding marker $w(x)$ matching the bulk invariants of the three corresponding infinite-size systems, up to fluctuations due to disorder. Here we consider a Kitaev chain (see main text) with total length $L=400$, parameters $J_1=1$, $J_2=0.65$, $\delta J_1=0.02$, $\delta J_2=\delta \mu =0$, and $\mu = 0.2$, $0.8$ or $3.6$ (from left to right).}
\end{figure}

Topological insulators and superconductors are phases of matter characterised by the exact quantisation of macroscopic observables and the appearance of edge states at the boundary of open systems~\cite{Hasan2010}. Such peculiar edge states include condensed-matter realisations of Majorana bound stated and unidirectional edge states, which are particularly robust against disorder and local perturbations. This makes them particularly appealing to engineer devices such as qubits, quantum channels \cite{Dlaska2017}, and eventually quantum computers~\cite{Nayak2008}. In a quantum device, several components in different topological phases can be brought together; see Fig.~\ref{fig_winding_marker}. 
Therefore, it is convenient to have a means of \emph{locally} discriminating between different topological phases. To this end, we use a topological winding marker originally introduced to study topological Anderson insulators~\cite{Meier2018} and defined in analogy with the two-dimensional Chern marker~\cite{Bianco2011,Caio2018}. This quantity locally distinguishes topological phases of one-dimensional systems with chiral symmetry \cite{Meier2018}. This is in contrast both with the global approach of standard topological invariants that are only defined for infinite systems~\cite{Hasan2010}, and with approaches based on scattering matrices~\cite{Akhmerov2011,Fulga2011,Fulga2012,Beenakker2015} that fundamentally characterise an interface. Although attempting a direct measurement of the winding marker in solid-state systems would raise numerous challenges, we will show that it can be related to readily available experimental data.

In one-dimensional topological insulators and superconductors, the local density of states (LDOS) can be obtained from the tunnelling differential conductance, observed by scanning tunnelling microscopy (STM) \cite{NadjPerge2014,Ruby2015,Pawlak2016,Feldman2016,Jeon2017}, or with more elaborate setups \cite{Zhang2018b}. STM provides a relatively non-invasive measurement of the LDOS as it does not require the deposition of contacts; this might be relevant for the non-destructive testing of topological devices, e.g. to assess whether a manufactured sample is in the expected topological phase. Although the LDOS of a system without edges does not carry information about its topology, the edge states will appear in the LDOS of a finite size system. However, these may be obscured by the presence of disorder in the sample. Moreover, STM measurements only allow to access the LDOS up to an unknown prefactor \cite{Chen2007}. The relation between the measured LDOS and the winding marker can therefore be subtle, even in the absence of disorder, but we shall see that it can be inferred using supervised machine learning.

Machine learning techniques are increasingly used in physics \cite{Mehta2018,Carleo2017,Carrasquilla2017,Baireuther2018,Baireuther2018b}. In particular, several works applied machine learning to study topological phases, mostly focusing on their classification from numerically accessible quantities such as entanglement spectra \cite{vanNieuwenburg2017}, density matrices \cite{Carvalho2018}, Hamiltonians \cite{Zhang2018,RodriguezNieva2018} or their eigenvectors \cite{Huembeli2018,Holanda2019}, loops of two-point correlation functions \cite{Zhang2017}, the density profile of quantum walks \cite{Ming2018}, the local density of a single state \cite{Ohtsuki2016, Ohtsuki2017, Araki2018}, and its disorder-averaged version \cite{Yoshioka2018}. In cold-atom systems, direct measurements of the topological invariants can be performed \cite{Atala2013,Aidelsburger2014,Jotzu2014,Meier2018} and nonetheless artificial neural networks have proven useful to identify topological phases from experimental momentum distributions \cite{Rem2018}. In solid-state systems however, the issue of determining the topological nature of a given sample from experimentally accessible data remains open.

In this article, we show that the winding marker in the centre of a finite size sample can be predicted from a measurement of the LDOS of the whole sample, by using supervised machine learning. Beside being able to distinguish trivial from topological phases, our method also discriminates between topological phases with distinct integer invariants. This is a non-trivial task as the simple counting of states is not available via STM measurement. Our method is of particular interest for unconventional superconductors, such as \ce{Sr2RuO4} \cite{Scaffidi2015,Kallin2016} and one-dimensional~\cite{Sahlberg2017} and two-dimensional \cite{Rontynen2015} Shiba lattices, where large values of topological invariants are predicted. In those systems, the experimental determination of the number of edge states is highly challenging due to the lack of easily accessible electrical transport signatures and because of the difficulties in the accurate measurement of the quantised thermal conductance at low temperatures~\cite{Jezouin2013,Banerjee2017,Banerjee2018,Kasahara2018}.

\vspace{0.5cm}
\paragraphTitle{Local winding marker}
Topological phases of matter are characterised by quantised topological invariants, typically defined globally for infinite systems. These invariants often manifest themselves in the response function of the ground state to an appropriate gauge field \cite{Qi2008,Ludwig2015}. Hence, we can expect them to correspond to reasonably localised quantities in real space: topological invariants can be recast in terms of the Fermi projector on the ground state, which is \emph{nearsighted} for gapped systems~\cite{Aizenman1998,Prodan2005,Bianco2011}. As first noticed by Bianco and Resta \cite{Bianco2011} in the case of the anomalous Hall effect, this enables a \emph{local} quantity closely related to the topological invariant to exist. 

In this work, we focus on one-dimensional systems with chiral symmetry, such as polyacetylene~\cite{Su1979} or the Kitaev chain~\cite{Kitaev2001}. The chiral symmetry is realised by a unitary operator $\hat\Gamma$ which anticommutes with the Hamiltonian $\hat H$. This class of topological systems is characterised by an integer-valued invariant called winding number, defined in momentum space as~\cite{Chiu2016}
\begin{equation}
	 w = \frac{\ii}{2\pi} \int_{\text{BZ}} \tr(\hat Q_{\text{LR}}(k) \partial_k \hat Q_{\text{RL}}(k)) \,\dd k.
\end{equation}
Here, $\hat Q_{\text{RL(LR)}} = \hat P_{\text{R(L)}} \hat Q \hat P_{\text{L(R)}}$, where $\hat P_{\text{R(L)}} = (\hat \Id \pm \hat \Gamma)/2$, $\hat Q = \Id - 2 \hat P_{\text{F}}$, and $\hat P_{\text{F}}$ is the projector on the states below the Fermi level.
While convenient, translation invariance is not necessary to define the winding number. In an infinite system, it can be defined as the trace per unit volume~\cite{MondragonShem2014,Song2014,Rakovszky2017}
\begin{equation}
	\label{real_space_winding_number}
	w = \frac{\ii}{2\pi} \Tr_{\text{vol}} (\hat Q_{\text{LR}}  [2 \pi \ii \hat X, \hat Q_{\text{RL}}])
\end{equation}
where $\hat X$ is the position operator. In particular, this real-space formulation applies to disordered systems. 
The topological invariant in \eqref{real_space_winding_number} is a \emph{global} quantity, quantised even at strong disorder \cite{MondragonShem2014,Song2014,Prodan2016}. 

\begin{figure}[b]
	\includegraphics{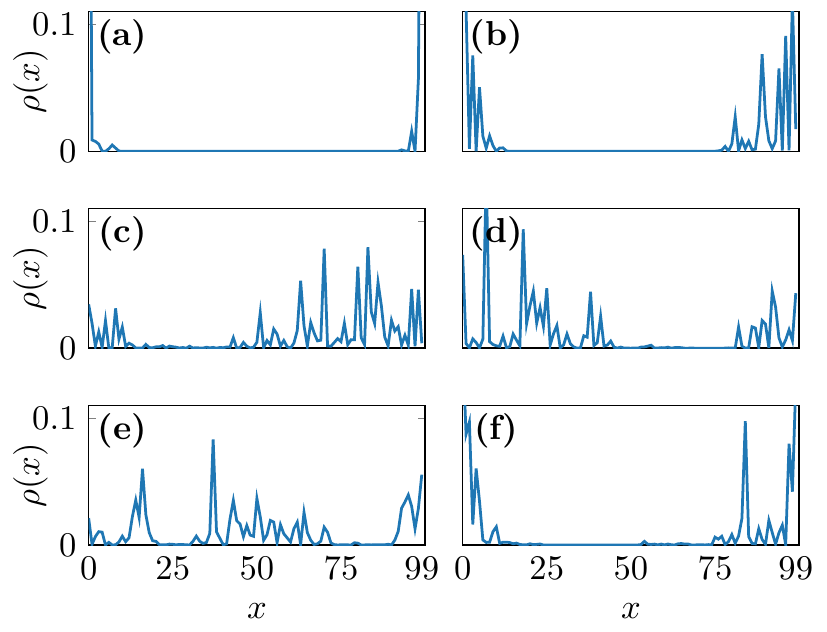}
	\caption{\label{fig:ldos_examples}\strong{Local density of states of topological and trivial samples.} (a-f) Examples of normalised local densities of states $\rho(x)$ for several disordered Kitaev chains of length $L=100$, corresponding to winding numbers $\bar w\simeq0,-1$, and $-2$. It is relatively easy to guess that (a), (b) and (f) are in a topological phase, as there are clear edge states on the boundaries, and nothing in the bulk. However, it is not obvious that (a) and (b) have $\bar w\simeq -1$, while (f) has $\bar w\simeq -2$.
	It is even harder to identify the topology in the cases (c-e), because of the many peaks in the LDOS due to the disorder. It turns out that (c) corresponds to $\bar w\simeq-1$, while (d) and (e) correspond to $\bar w\simeq 0$.}
\end{figure}

In analogy with the Chern marker for two-dimensional systems introduced by Bianco and Resta~\cite{Bianco2011}, a local winding marker was introduced by Meier et al.~\cite{Meier2018} as
\begin{equation}
	w(x) = \frac{-1}{V_{\text{uc}}} \sum_{\alpha} \braket{x, \alpha \mid \hat Q_{\text{LR}} [\hat X, \hat Q_{\text{RL}}] \mid x, \alpha},
\end{equation}
where $x$ is the position along the chain, $\alpha$ labels the degrees of freedom in the unit cell of the Bravais lattice, and $V_{\text{uc}}$ is the volume of the unit cell. While we focus here on one-dimensional systems, the same construction is available in all odd space dimensions.

The local winding marker can be computed for the experimentally relevant case of disordered finite-size systems with open boundaries and, notably, for composite systems.
In Fig.~\ref{fig_winding_marker}, a chain is divided in three regions, each with different parameters of the Hamiltonian. Infinite-size systems with the corresponding parameters would have three different winding numbers. Away from the interfaces, the winding marker displays plateaux at the corresponding values, up to fluctuations due to disorder.

\vspace{0.5cm}
\paragraphTitle{Neural network assisted measurement}
In order to infer the value of the winding marker from accessible experimental data, we use supervised machine learning in the form of a feedforward neural network.
The spatially resolved density of the states close to the Fermi energy can be measured in STM experiments, as discussed~\cite{Chevallier2016} and observed~\cite{NadjPerge2014,Ruby2015,Pawlak2016,Feldman2016,Jeon2017} in the context of one-dimensional topological systems. However, there is little control on the number of states involved in the measurement when the system is disordered, and the LDOS can be measured only up to an unknown prefactor. In order to model such a measurement, we use as input of our neural network the LDOS corresponding to an energy window $\mathcal{E}$ of size $\delta E$ centred at the Fermi energy
\begin{equation}
	\rho(x) = \frac{1}{N} \sum_{\alpha,E_i\in \mathcal{E}} |\braket{x, \alpha \mid \psi_{i}}|^{2}.
\end{equation}
Here, the sum runs over the internal degrees of freedom~$\alpha$, and over the eigenstates $\ket{\psi_{i}}$ of the Hamiltonian $\hat H$ with energies $E_i\in\mathcal{E}$. Besides, $N$ is a normalisation constant ensuring that $\int \rho(x) \dd x = 1$, so that the neural network does not simply count the total number of states in the window. In our numerical calculations, we set $\delta E$ to be \SI{1}{\percent} of the bandwidth. In Fig.~\ref{fig:ldos_examples}, we show some examples of normalised LDOS drawn from the dataset used to train our neural network. A visual analysis reveals no obvious connection between the shape of the LDOS and the number of topological edge states.

Although the winding marker is defined locally, we are interested in its value in the bulk of the system. Away from the sample boundaries, the winding marker $w(x)$ corresponds to the topological invariant $w$, up to fluctuations due to disorder. To remove these, we label each item in the training set of the neural network with the average $\bar w$ of $w(x)$ over a region of size $\ell=L/3$ in the centre of the sample of size $L$. The neural network is then trained to predict $\bar w$ from a normalised LDOS. Details about the architecture, implementation, and $K$-fold cross-validation training and testing of the feedforward neural network are discussed in the Supplemental Material.

\begin{figure}
 \includegraphics[width=3.2in]{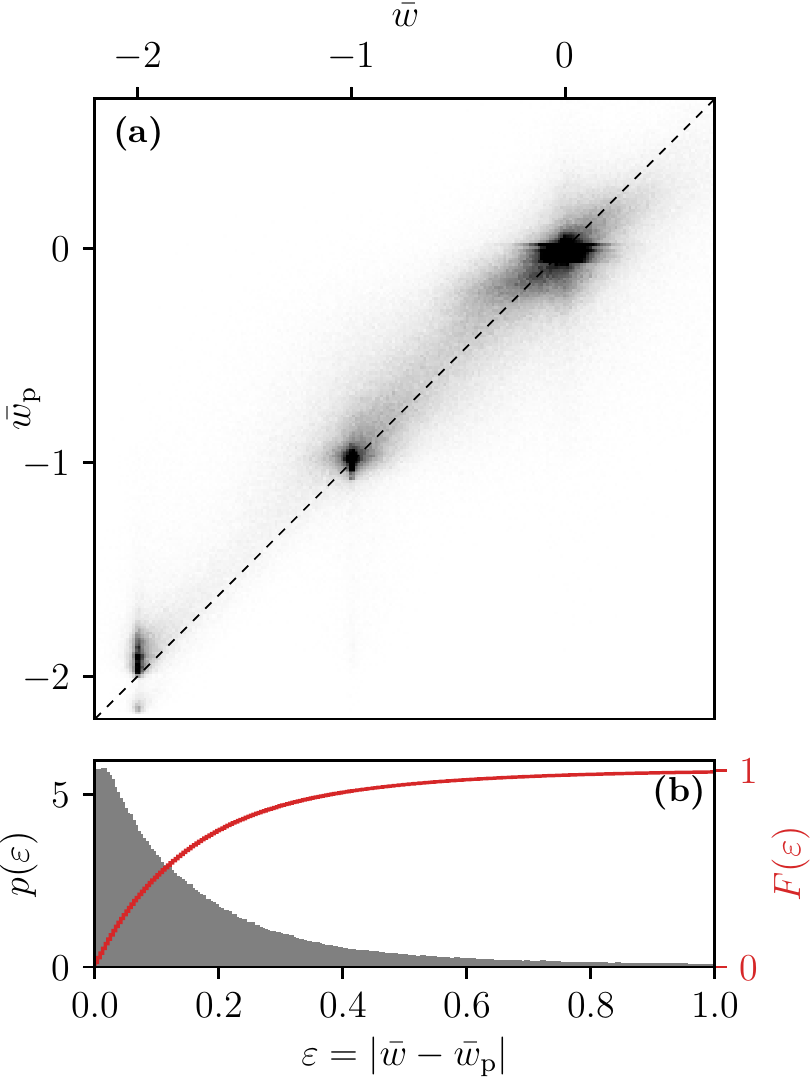}
 \caption{\label{fig:predictions} \strong{Neural network prediction of the winding marker.}
 (a) Normalised distribution $p(\bar w,\bar w_p)$ of the predicted winding marker $\bar{w}_{\text{p}}$ with respect to the actual averaged winding marker $\bar{w}$. Most of the data concentrate in spots $(n,n)$ on the diagonal (dashed line), corresponding to integer values $n=0,-1,-2$ of the winding marker. Data outside of these spots mostly correspond to parameters of the Hamiltonian close to phase transitions. The spot at $(0,0)$ is less sharp than the others, as many data points correspond to phases were the topology is made trivial by disorder. (b) Normalised distribution $p(\varepsilon)$ and cumulative distribution function $F(\varepsilon)=\int_0^\varepsilon p(\varepsilon^\prime) \dd\varepsilon^\prime$ of the error $\varepsilon=|\bar{w}-\bar{w}_{\text{p}}|$. The corresponding mean absolute error is \num{0.18}.}
\end{figure}%
\vspace{0.5cm}
\paragraphTitle{Results}
\noindent In this work, we focus on the disordered Kitaev chain~\cite{Kitaev2001}, where we include next to nearest neighbours hoppings in order to explore the $w=-2$ phase, in addition to the usual $w=0,-1$ phases. For simplicity, we assume the hopping terms to be equal to the superconducting pairings, and consider the Hamiltonian $\hat H=\sum_{x,x'} H_{x,x'} \ket{x} \bra{x'}$, where $H_{x,x} = \mu(x) \tau_z$, $H_{x,x+1} = H_{x,x-1}^\dagger= J_1(x) (\tau_z + \ii \tau_y)$, $H_{x,x+2} = H_{x,x-2}^\dagger= J_2(x) (\tau_z + \ii \tau_y)$, where $\tau_{\nu}$ are Pauli matrices in particle-hole space. Here, we consider uncorrelated disorder where $\mu(x) \sim \mathcal{N}(\mu, \delta \mu)$ are independent and identically distributed random variables following a normal distribution with mean $\mu$ and standard deviation $\delta \mu$, and similarly for $J_{1(2)}(x) \sim \mathcal{N}(J_{1(2)}, \delta J_{1(2)})$.
This Hamiltonian has both particle-hole and chiral symmetries.
In a generic superconducting system, only particle-hole symmetry is present, and our method might be adapted to assess the corresponding $\ZZ_2$ topology. 
Here, we focus on the more delicate situation where several topologically non-trivial phases have to be distinguished.

The dataset for the training and testing of the neural network consists of \num{906250} tuples $(\rho(x),\bar w)$ obtained by randomly drawing the parameters $J_{1(2)}$, $\delta\mu$ and $\delta J_{1(2)}$ uniformly from the interval $[0,1)$ and $\mu$ from $[0,2)$. In Fig.~\ref{fig:predictions}, we show the two-dimensional distribution of the predicted winding marker $\bar{w}_\text{p}$ with respect to the actual (directly calculated) averaged winding marker $\bar w$. For a perfect prediction, all the data points should lie on the diagonal; and, for a perfectly quantised winding marker, all the data should concentrate at the points $(n,n)$ for $n = 0,-1,-2$. In Fig.~\ref{fig:predictions}(a), three spots are indeed clearly visible, and their finite width is due to the presence of disorder. The normalised distribution of the error $|\bar w - \bar w_\text{p}|$, in Fig.~\ref{fig:predictions}(b), shows the accuracy of the predictions. For our trained neural network, we obtain a root mean squared error $\text{RMSE} = \num{0.275 \pm 0.003}$. The tail in the distribution of errors in Fig.~\ref{fig:predictions}(b) corresponds to the subtle vertical features in Fig.~\ref{fig:predictions}(a), where the error is larger. 
In order to test the scalability of our approach, we consider a system twice as big, with length $L=200$. We obtain a similar $\text{RMSE} = \num{0.282 \pm 0.002}$ using a dataset composed of \num{2052096} tuples $(\rho(x),\bar{w})$. The influence of the size of the dataset on the MSE is discussed in the Supplemental Material.

We expect the network to recognise features associated with the topological edge states, and not inessential features specific to the system. To verify this hypothesis, we train and test the same neural network using as input the LDOS of a sample of length $L=\num{200}$ restricted to the $\num{100}$ central sites. As expected, the network trained in this way loses any predictive ability; see Supplemental Material. 
In Fig.~\ref{fig:phase_diagram}, we show a slice of the phase diagram of the disordered Kitaev chain, comparing the values of (a) the predicted winding marker $\bar{w}_{\text{p}}$ and (b) the spatially averaged winding marker $\bar{w}$ over a range of parameters, for a single disorder realisation.
Further, in panels (c) and (d) of Fig.~\ref{fig:phase_diagram}, we show their average over \num{50} disorder realisations.  The remarkable agreement between the actual and predicted winding markers illustrate the accuracy of the network in parameter space, even for large disorder.

\medskip

\begin{figure}
 \includegraphics[width=3.2in]{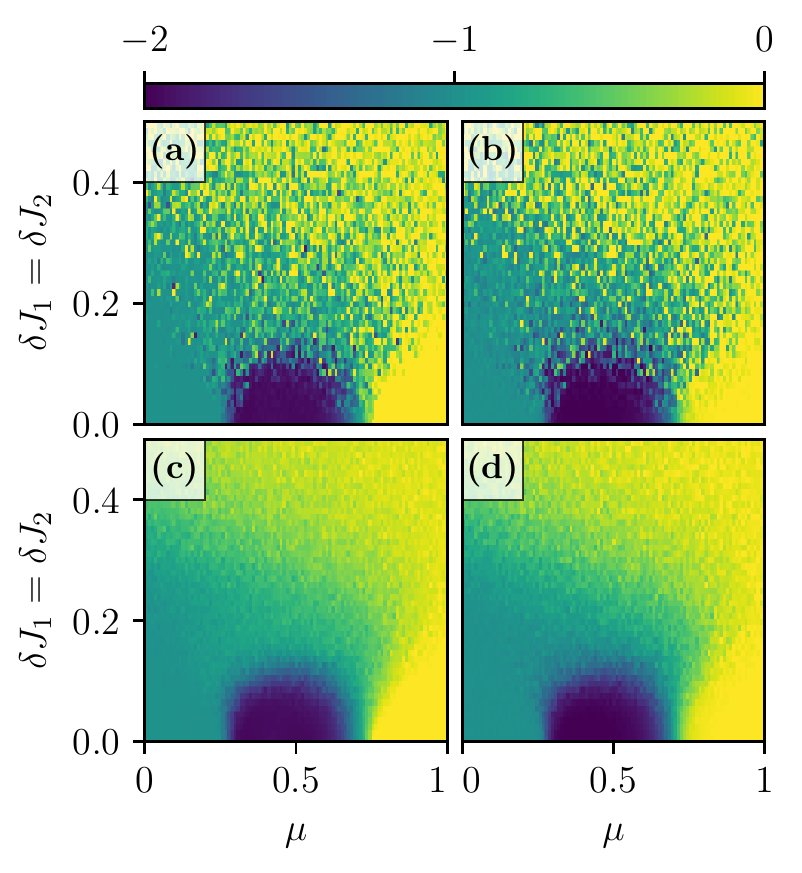}
 \caption{\label{fig:phase_diagram} \strong{Predicted and reference phase diagrams.}
 (a) Predicted winding marker $\bar{w}_{\text{p}}$ for one disorder realisation.
 (b) Actual winding marker $\bar{w}$ for one disorder realisation.
 (c) Predicted winding marker $\braket{\bar{w}_{\text{p}}}_{\text{dis.}}$ averaged over \num{50} disorder realisations.
 (d) Actual winding marker $\braket{\bar{w}}_{\text{dis.}}$ averaged over \num{50} disorder realisations.
 This slice of the phase diagram, corresponding to different values of the onsite potential $\mu$ and of the disorder amplitudes $\delta J_1 = \delta J_2$, is computed for a system of length $L=\num{100}$, with $J_1 = \num{0.5}$ and $J_2=\num{0.375}$. }
\end{figure}

So far, we have used a neural network to infer the bulk topology of a homogeneous finite-size chain from its LDOS. Further, we can take advantage of the local character of the winding marker by applying our method to a composite chain. As a proof of principle, we focus on the simplest example where the left and right halves of a one-dimensional chain of size $L_{\text{c}}$ are potentially different. More precisely, both the average values and the standard deviations of the parameters $\mu(x)$ and $J_{1}(x)$ are independently chosen for the left and the right of the chain. For simplicity, $J_{2}(x)$ is set to identically vanish throughout the chain, which implies that the winding number can be either $-1$ or $0$, on each side. The same procedure as before is then applied: the LDOS of the entire chain is used as the input of a feedforward neural network, which is trained using as labels the averages $\bar{w}_{\text{L}}$ and $\bar{w}_{\text{R}}$ of the winding marker over regions of size $\ell=L_{\text{c}}/6$ centred at $L_{\text{c}}/4$ and $3 L_{\text{c}}/4$. The network outputs the predicted values of the averaged winding markers $\bar{w}_{\text{p,L}}$ and $\bar{w}_{\text{p,R}}$. Here, $\text{L}$ and $\text{R}$ respectively label the left and right sides of the chain.

\begin{figure}
 \includegraphics[width=3.2in]{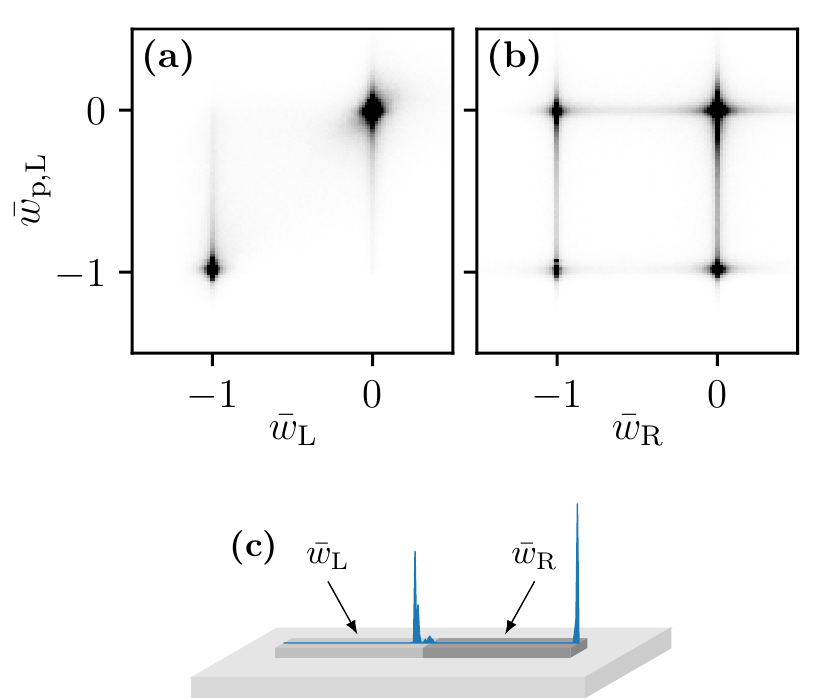}
 \caption{\label{fig:distribution_composite} \strong{Neural network prediction for a composite system.}
 (a) Normalised marginal distribution $p(\bar{w}_{\text{L}},\bar{w}_{\text{p,L}})$
 of the predicted winding marker $\bar{w}_{\text{p,L}}$ for the left half of the chain with respect to the actual averaged winding marker $\bar{w}_{\text{L}}$ for the left (same) half. 
 (b) Normalised marginal distribution $p(\bar{w}_{\text{R}},\bar{w}_{\text{p,L}})$
 of the predicted winding marker $\bar{w}_{\text{p,L}}$ for the left half of the chain with respect to the actual averaged winding marker $\bar{w}_{\text{R}}$ for the right (other) half.
 (c) Sketch of the composite system, with an example of LDOS in blue; in this case $\bar{w}_{\text{L}} \simeq \num{0}$ while $\bar{w}_{\text{R}} \simeq \num{-1}$.
 The four-dimensional histogram from which the marginals are obtained is computed with \num{180} bins for each dimension.}
\end{figure}

In Fig.~\ref{fig:distribution_composite}, we show the two-dimensional marginal distributions (a) $p(\bar{w}_{\text{L}},\bar{w}_{\text{p,L}})$ and (b) $p(\bar{w}_{\text{R}},\bar{w}_{\text{p,L}})$, for a chain of length $L_\text{c}=200$, and \num{1537536} tuples $(\rho(x), (\bar{w}_\text{L},\bar{w}_{\text{R}}))$ in the dataset. The neural network is identical as for Fig.~\ref{fig:predictions}, except for the output layer which now includes two units.
As expected, Fig.~\ref{fig:distribution_composite}(a) resembles Fig.~\ref{fig:predictions}(a) while Fig.~\ref{fig:distribution_composite}(b) shows the lack of any meaningful correlation between the predicted value of the left half and the actual value of the right half. The marginals $p(\bar{w}_{\text{R}},\bar{w}_{\text{p,R}})$ and $p(\bar{w}_{\text{L}},\bar{w}_{\text{p,R}})$ (not shown) have identical features. For the trained neural network, we obtain a $\text{RMSE} = \num{0.289 \pm 0.001}$.

\vspace{0.5cm}
\paragraphTitle{Discussion}\label{sec:discussion}
In this article, we have shown that machine learning techniques allow to infer the average of the local winding marker from the experimentally accessible local density of states. This enables to characterize the topology of finite-size and, in particular, composite one-dimensional chiral systems. Crucially, not only are we able to distinguish topological from non-topological phases but we can also discriminate between topological phases with different invariants. Our approach is as non-invasive as possible, and is suitable even for systems where direct transport measurements cannot be used.

While the winding marker is a genuinely local quantity, the neural network fundamentally recognises interfaces, as it relies upon the appearance of topological edge states in the local density of states. This is reminiscent of the scattering matrix description of topological systems. Although the neural network predicts spatially averaged values of the winding marker, we have shown that it can locally predict the topology of adjacent regions in a composite system.  

Here, we focused on a proof of concept where the LDOS and the topological marker are determined from a specific family of tight-binding Hamiltonians.
When a larger family of Hamiltonians is considered, e.g. including more degrees of freedom and longer range hoppings, a larger network and training set might be required to maintain the same level of accuracy. For example, we expect our approach to distinguish even larger values of the topological invariant, possibly at a cost of an increased size of the neural network. In most experimental setups, the parameter space of the Hamiltonians is strongly constrained by the symmetries and the locality. Therefore, we expect that for each setup, it is possible to tailor and train a network which can efficiently identify the distinct topological phases.
We can also wonder if a finite training set is enough to learn a general rule, allowing the predictions to remain accurate even when the Hamiltonians are generated from larger and larger subsets of the whole set of class BDI Hamiltonians. This question goes beyond the scope of this work, but is highly interesting from a fundamental point of view.
Future directions for research also include extensions to the Chern marker for two-dimensional systems, as well as to other local topological markers.

\paragraphTitle{Acknowledgments}
We thank J. Tworzyd\l{}o and C. Beenakker for fruitful discussions.
This research was supported by the Netherlands Organisation for Scientific Research (NWO/OCW) as part of the Frontiers of Nanoscience (NanoFront) program, by an ERC Synergy Grant, and by the Foundation for Polish Science through the IRA Programme, co-financed by EU within SG OP.

\bibliography{bibliography}

\onecolumngrid
\clearpage

\appendix
\twocolumngrid
\section{Neural network architecture}

The input $\vec{\rho}$ of our model is an array of nonnegative real numbers of fixed length $N_{\text{in}} = \num{100}$ representing the normalised local density of states (LDOS) of the finite system close to the Fermi energy; its output is the predicted winding number $\bar{w}_{\text{p}}$, a single real number.
For this regression task, we employ a feedforward artifical neural network composed of \num{3} hidden layers. 
Each hidden layer $\vec{h}^{i}$, with $i\in\{1,2,3\}$, contains \num{128} rectified linear units (ReLU) to provide non-linearity, followed by a batch normalisation (BN) \cite{ioffe2015batch} to speed up and stabilise the training by reduction of the internal covariance shift.
The output layer $\mathcal{L}^4$ is a single unit with linear activation, which corresponds to a linear mapping from the last hidden layer.
To regularise the model and thus prevent overfitting, during training we apply dropout~\cite{Srivastava2014} to the output of the last hidden layer with a dropout probability of \num{0.5}.
The weights and biases of the network are fitted using the Adam optimizer~\cite{Kingma2014}, with an learning rate of \num{0.001}.
We use the mean squared error (MSE) as the loss function to train the parameters of the neural network, as it provides both a suitable metric for the regression problem and can be used in backpropagation, being differentiable with respect to the network weights.
The implementation is done with the Keras package~\cite{Keras}, using TensorFlow~\cite{Tensorflow} as backend. 

The network is formally a function space of maps parametrised by a set of weights matrices $M^n$ and bias vectors $b^n$ that can be expressed as
\begin{align}
 \bar{w}_{\text{p}} = [\mathcal{L}^4 \circ \vec{h}^3 \circ \vec{h}^2 \circ \vec{h}^1](\vec{\rho})
\end{align}
where
\begin{align}
  h^n_i &= \text{BN} \circ \mathcal{F} \circ \mathcal{L}^n_i, \\
  \mathcal L^n_i(\vec x) &= M_{ij}^n \, x_j + b_i^n, \\
  \mathcal{F}(y) &= \max(0, y).
\end{align}
The transformation $\vec{x} \mapsto \text{BN}(\vec{x})$ is described in Algorithm~1 of Ref.~\cite{ioffe2015batch}. 
It is parametrised by the feature-wise mean and variance values, which are fitted during training.

For the case of a bipartite composite system, the output of the model is a vector $(\bar{w}_{\text{p,L}}, \bar{w}_{\text{p,R}})$. The only change to the neural network architecture is that in this case, the output layer $\mathcal{L}^4$ is now composed of two neurons with linear activation.

\medskip

The model architecture is obtained by evaluating the MSE of different architectures on a fixed test set when trained on a fixed training set; a separate validation set is used for interrupting the training when the MSE on it is no longer decreasing. The contending ML models were AlexNet-like convolutional neural networks, boosted trees, and support vector machine with linear kernel, but a feedforward neural network largely outperformed all of them. Once the general architecture of the model is found, the same strategy of MSE evaluation is used again to choose the hyperparameters of the network (e.g. number of hidden neurons, number of layers, dropout ratio). 

After the architecture of the model is determined, a new dataset is generated to train and test the network. The presented results are obtained by $K$-fold cross-validation with $K = \num{10}$, where the whole dataset is randomly split in $K$ \enquote{folds} each containing the same fraction of data. One at a time, each fold is used for testing whereas the other $K - 1$ are used for training. This allows us to estimate the expected error of our method as well as the uncertainty on this expected error, by computing the the average and standard deviation of the MSE for all the folds in the cross-validation.

\section{Kitaev model with second nearest neighbours}

The Kitaev Hamiltonian with second nearest neighbours reads
\begin{equation}
\begin{split}
	H &= - \sum_{x} \mu(x) \tau_{z} \ket{x}\bra{x} \\
	 &- \sum_{x} (t_1(x) \tau_{z}  + \ii \Delta_1(x) \tau_{y} + \ii \eta_1(x) \tau_{x}) \ket{x}\bra{x+1} + \text{h.c.} \\
	&- \sum_{x} (t_2(x) \tau_{z}  + \ii \Delta_2(x) \tau_{y} + \ii \eta_2(x) \tau_{x}) \ket{x}\bra{x+2} + \text{h.c.}
\end{split}
\end{equation}
The terms proportional to $\eta_1$ and $\eta_2$ break chiral symmetry $\Gamma = \tau_{x}$ and time-reversal symmetry $\Theta = \mathcal{K}$ (complex conjugation), but preserve particle-hole symmetry $\Xi = \Gamma \Theta$. As such, they collapse the $\ZZ$ invariant to a $\ZZ_2$ invariant when present.

In the main text, we consider $t_{1(2)}(x) = \Delta_{1(2)}(x) =: J_{1(2)}(x)$ for simplicity, and set $\eta_1(x)=\eta_2(x)=0$ to preserve chiral symmetry. We mainly consider disordered homogeneous systems, where $\mu(x) \sim \mathcal{N}(\mu, \delta \mu)$
are independent and identically distributed random variables following a normal distribution with mean $\mu$ and standard deviation $\delta \mu$, and similarly for $J_{1(2)}(x) \sim \mathcal{N}(J_{1(2)}, \delta J_{1(2)}) $.

When all parameters are uniform in space, the system is translation invariant and one can block-diagonalize it in Bloch representation, where the Bloch Hamiltonian is
\begin{equation}
\begin{split}
	H(k) = &(-\mu - 2 t_1 \cos(k) - 2 t_2 \cos(2 k)) \tau_{z} \\
	+ &(2 \Delta_1 \sin(k) + 2 \Delta_2 \sin(2 k)) \tau_{y} \\
	+ &(2 \eta_1 \sin(k) + 2 \eta_2 \sin(2 k)) \tau_{x}.
\end{split}
\end{equation}

\section{Additional data}

\begin{figure}
 \includegraphics[width=3.2in]{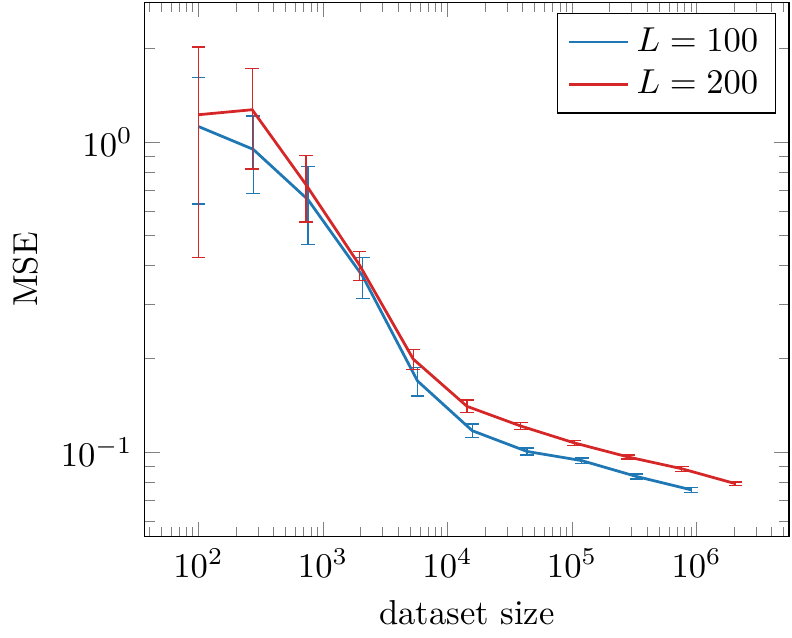}
 \caption{\label{fig:mse_dataset_size} \strong{Mean squared error for different dataset sizes.}
 A series of subsets with sizes evenly distributed on a logarithmic scale are randomly drawn from the main dataset. The MSE is computed for each subset, and represented on a log-log plot. The same is done for $L=\num{100}$ and for $L=\num{200}$. Error bars represent the estimated uncertainty on the MSE.
 }
\end{figure}

\begin{figure}
 \includegraphics[width=3.2in]{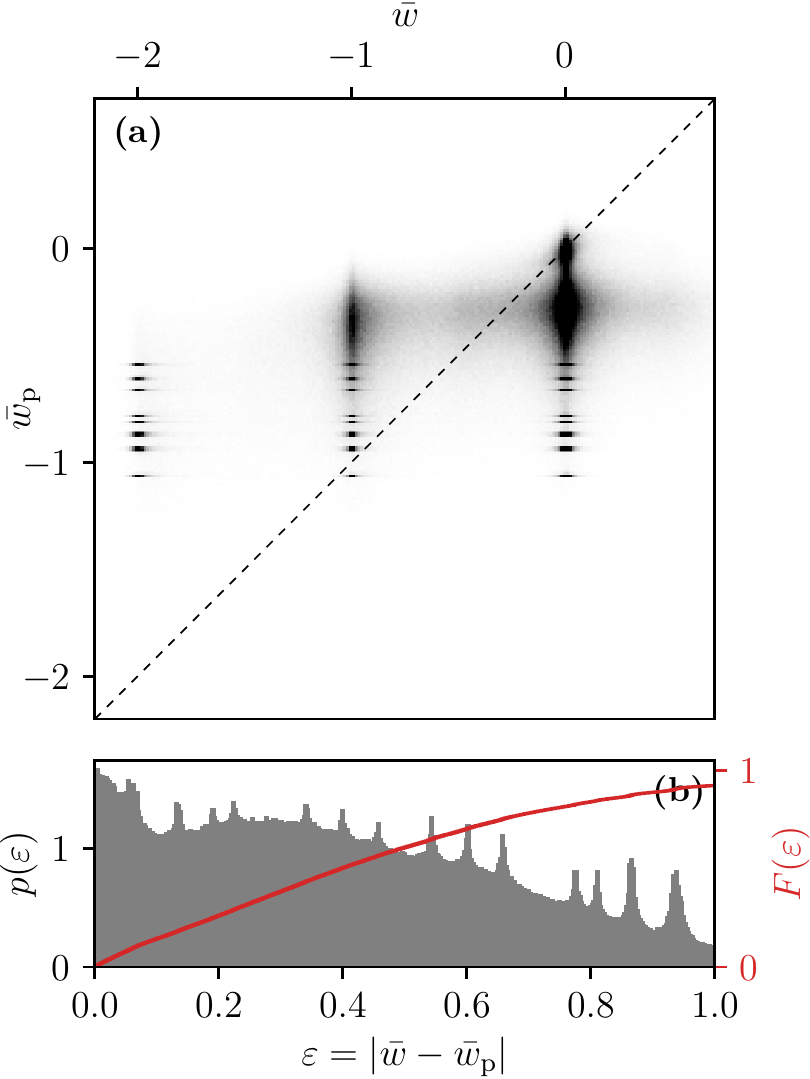}
 \caption{\label{fig:predictions_from_bulk} \strong{Predictions from a bulk LDOS.}
 (a) Normalised distribution $p(\bar w,\bar w_p)$ of the predicted winding marker $\bar{w}_{\text{p}}$ with respect to the actual averaged winding marker $\bar{w}$, for a system of size $L=\num{200}$ where we only consider the LDOS on the $\num{100}$ central sites (the middle of the chain).
 The dashed line corresponds to $\bar{w}=\bar{w}_{\text{p}}$.
 There is hardly any correlation between the actual averaged winding marker and the prediction.
 (b) Normalised distribution $p(\varepsilon)$ and cumulative distribution function $F(\varepsilon)=\int_0^\varepsilon p(\varepsilon^\prime) \dd\varepsilon^\prime$ of the error $\varepsilon=|\bar{w}-\bar{w}_{\text{p}}|$. The corresponding mean absolute error is \num{0.46} (the RMSE is \num{0.275}).}
\end{figure}

The influence of the size of the dataset on the performance of the model is illustrated in Fig.~\ref{fig:mse_dataset_size}. 
A series of subsets with sizes evenly distributed on a logarithmic scale are randomly drawn from the main dataset.
For each subset, the MSE and its uncertainty are estimated using the $K$-fold cross-validation training and testing procedure discussed in the section \emph{Neural network architecture}. The MSE quickly decreases up to a dataset size of roughly \num{1e4}. For larger datasets, a slower decrease of the MSE compatible with a linear behaviour on logarithmic scale is observed. The uncertainty on the MSE, which represents the variability for different folds in the $K$-fold procedure, is notably larger in the first part. This might be due to an inadequate sampling of the parameter space for small datasets, due to an insufficient number of data points. The same behaviour is observed both for chains of length $L=\num{100}$ and $L=\num{200}$; in the second part of the graph, the curves for both lengths appear to be parallel with a constant offset, up to the uncertainties.

\medskip

We presume that the machine learning procedure distinguishes systems with different topological invariants from the existence and shape of the topologically protected edge states expected from the bulk-boundary correspondence principle. Hence, it should not be possible to learn anything from the LDOS of a bulk system. To verify this, we train and test the same neural network as in the main text, but using as input the LDOS of a sample of length $L=\num{200}$ where the LDOS is restricted to the $100$ central sites. In this region, the edge states have a vanishing contribution for most of the system parameters. The two-dimensional histogram and the distribution of the absolute error in Fig.~\ref{fig:predictions_from_bulk} show that the artificial neural network has lost any meaningful predictive ability. Correspondingly, the RMSE \num{0.568 \pm 0.006} is relatively large, although this value alone would not be sufficient to draw a conclusion. The origin of the peaks in the marginal distribution $p(\bar{w}_{\text{p}})$ is not clear, but is probably not physical; they may correspond to noise amplified by the artificial neural network.

\onecolumngrid

\end{document}